\documentclass{smarthepnote}

\usepackage[colorinlistoftodos]{todonotes}
\usepackage{placeins}

\title{The SMARTHEP European Training Network}
\author{The SMARTHEP Network}

\date{Preprint prepared \today}

\DocAuthors{L.~Bozianu \\ C.~Cocha~Toapaxi \\ P.Jawahar \\ M.~Olocco}
\DocEditors{J.~A.~Gooding}
\DocCoordinators{N/A}
\DeliverableNo{xxx-yyy-zzzz}
\draftversion{3.0}

\begin{document}
\maketitle

\begin{abstract}
    Synergies between MAchine learning, Real-Time analysis and Hybrid architectures for efficient Event Processing and decision-making (SMARTHEP) is a European Training Network, training a new generation of Early Stage Researchers (ESRs) to advance real-time decision-making, driving data-collection and analysis towards synonymity.

    SMARTHEP brings together scientists from major LHC collaborations at the frontiers of real-time analysis (RTA) and key specialists from computer science and industry. By solving concrete problems as a community, SMARTHEP will further the adoption of RTA techniques, enabling future High Energy Physics (HEP) discoveries and generating impact in industry.

    ESRs will contribute to European growth, leveraging their hands-on experience in machine learning and accelerators towards commercial deliverables in fields that can profit most from RTA, e.g., transport, manufacturing, and finance.

    This contribution presents the training and outreach plan for the network, and is intended as an opportunity for further collaboration and feedback from the CHEP community.
\end{abstract}

\vfill
\makereviewtable
\clearpage

\section{Introduction}
\label{intro}
The Synergies between MAchine learning, Real-Time analysis and Hybrid architectures for efficient Event Processing and decision making (SMARTHEP) European Training Network is an EU Horizon-funded training network, with a focus on the development of expertise in real-time analysis (RTA) techniques through applications to High Energy Physics (HEP) research and industry. The network centres around the training of 12 Early Stage Researchers (ESRs) between September 2022 and September 2025.

\section{SMARTHEP as a European Training Network}
\label{network}
As a European Training Network (ETN), the primary aim of the network is in training ESRs, whilst deepening synergies between HEP and industry. The network takes a novel approach to building such synergies, structuring each ESR position (a 3 year period of doctoral study) around academic and industrial secondments. To achieve this, the network is formed of a series of partnerships between universities, research institutes and organisations in industry, as listed in Table~\ref{partners}.

\begin{table}[h!]
    \centering
    \small
    \begin{tabular}{p{2.5cm}p{9.5cm}}
    \hline
    Category & Partners \\\hline
    Universities & Lund University, Sorbonne University (LPNHE \& LIP6), Technische Universit\"at Dortmund, University of Bologna, Universit{\'e} de Gen{\`e}ve, University of Heidelberg, University of Helsinki, University of Manchester, Universidade Santiago de Compostela (IGFAE), Vrije Universiteit Amsterdam \\\hline
    Research institutes & CERN, CNRS, NIKHEF  \\\hline
    Industry partners & IBM France, Lightbox, Point 8, University of Manchester Institute for Data Science and AI, Verizon Connect, Ximantis\\\hline
    \end{tabular}
    \caption{The member institutes and partners of the SMARTHEP network.}
    \label{partners}       
\end{table}

\begin{figure*}[h!]
    \centering
    \includegraphics[width=0.9\linewidth]{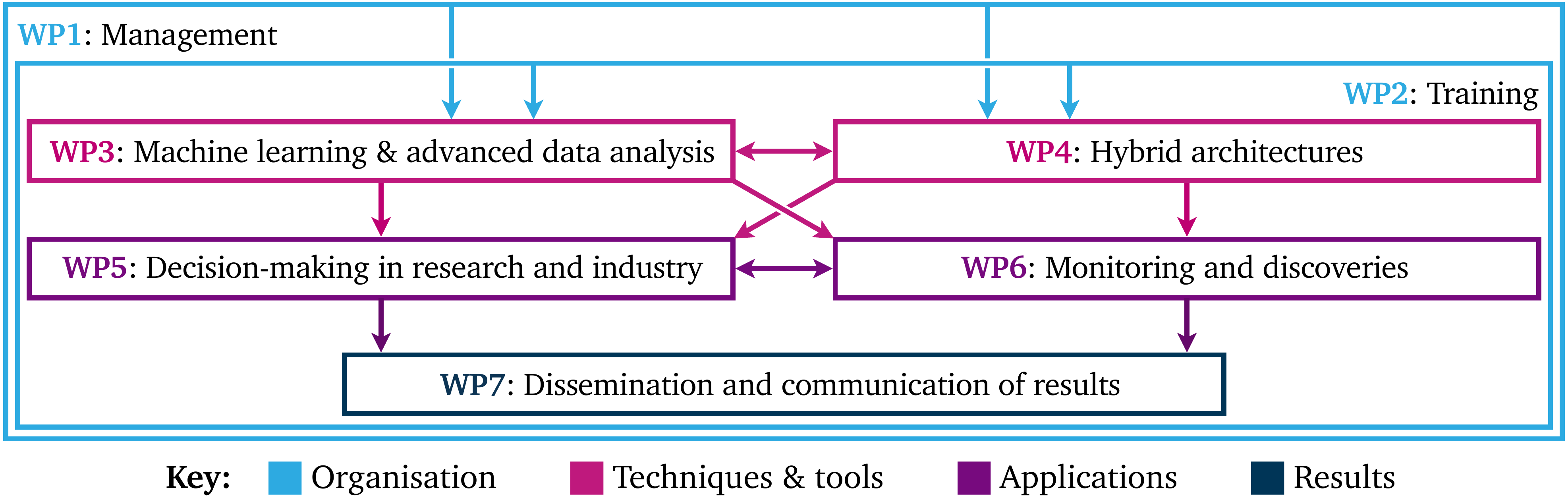}
    \caption{The structure of work packages within the SMARTHEP network. WP1 and WP2 define the organisation of the network; WP3 and WP4 introduce the techniques and tools of real-time analysis to the network; WP5 and WP6 use said techniques and tools to produce results for HEP and industry; WP7 makes these results available and promotes their wider use and adoption.}
    \label{network-diagram}
\end{figure*}

The network is formally structured as 7 Work Packages (WPs), laid out in Figure~\ref{network-diagram}. WP1 ``Management'' covers the management of the network by the Project Manager/Project Coordinator/Executive Board. WP2 ``Training'' sets out training and development of participants, by both network partners and external training providers. WP3 ``Machine learning \& advanced data analysis'' develops machine learning (ML) techniques for use in real-time environments. WP4 ``Hybrid architectures'' focuses upon the deployment of non-CPU computing architectures in acceleration of data processing. WP5 ``Decision-making in research and industry'', applies WP3 and WP4 to develop RTA-based decision-making technologies. WP6 ``Monitoring and discoveries'' also applies WP3 and WP4 to RTA approaches in data analysis. WP7 ``Dissemination and communication of results'' publishes and propagates the results produced in WP6 and WP7.

The network has a particular focus on physics at the Large Hadron Collider (LHC). Each ESR is thus affiliated to one of the four major experiments based at the LHC: ALICE, ATLAS, CMS and LHCb. The network duration coincides with Run 3 of the LHC (2022-2025), with many of the ESR projects framed in this context.

A unique feature of the network is the extensive cooperation between HEP and industry across all ESR positions. RTA approaches have seen significant adoption in industry in recent years, with many organisations turning to RTA as a means to handle the challenges of ``big data''. Industry has thus seen a swift growth in RTA expertise which can benefit symbiotically from such cooperation.

\section{Real-time analysis}
\label{rta}
\begin{figure*}[h!]
    \centering
    \includegraphics[width=\linewidth]{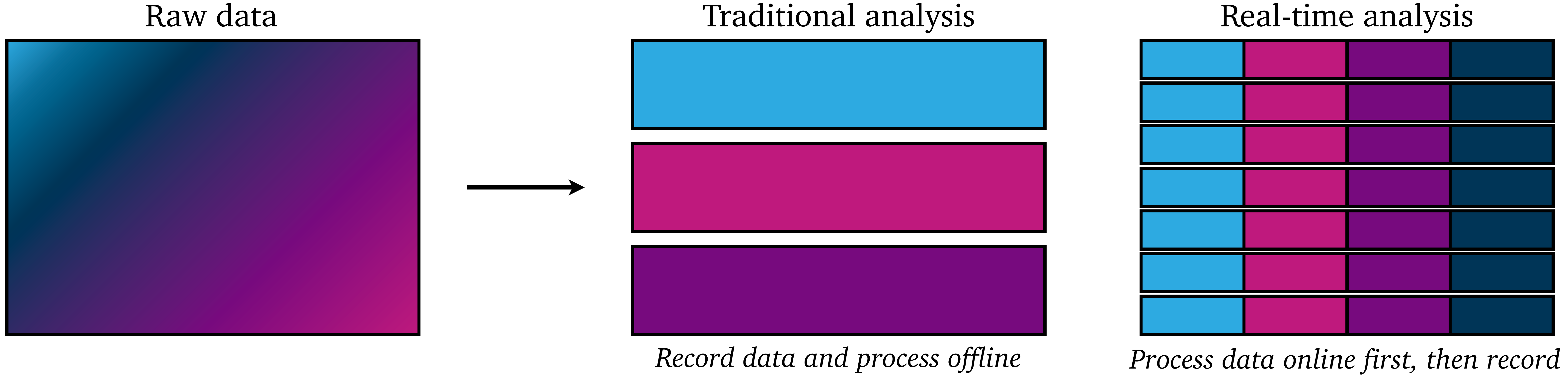}
    \caption{Traditional and RTA approaches to data processing. Traditional approaches rely on recording all data and processing this offline; in RTA, data is processed as it is produced, recording only the relevant portions, enabling greater volumes of processed data to be stored.}
    \label{rta-diagram}
\end{figure*}

HEP and industry share a common challenge—the rapid processing of large quantities of data~\cite{hu-big-data}. Recent advances in computing, in particular in the areas of machine learning and hybrid architectures, have enabled the possibility of processing data in real-time, i.e., as data is collected (also commonly referred to as ``online'' processing)~\cite{real-time-computing}. By processing data online, resources (computing power, storage space, energy, etc.) can be saved and further insights can be obtained from the data recorded, as shown in Figure~\ref{rta-diagram}.

RTA techniques have seen widespread adoption across HEP, in particular amongst trigger and data acquisition (TDAQ) systems. Since it is not possible to record a full detector readout and carry out event reconstruction at the LHC collision rate of {40}{ MHz}, triggers must be devised to select only those events relevant to the physics goals of an experiment. Such triggers conventionally consist of a hardware system making coarse decisions from partial detector readout, and a staged software trigger, applying gradually more finely-grained selections with increasingly more detailed reconstructions of events.

\subsection{Machine learning}
\label{machine-learning}
ML, a catch-all term for a family of techniques and technologies wherein algorithms are conditioned to analyse data, enables rapid decision-making and pattern recognition across a broad range of use cases~\cite{intro-ml}. In HEP, the adoption of ML began with classifiers for offline physics analysis, later widening to a variety of classification and pattern recognition/anomaly detection techniques for use online and offline~\cite{albertsson-ml}.

ML classifiers such as Boosted Decision Trees (BDTs) and Neural Networks (NNs) are commonly employed in HEP. The use of BDTs in signal selection for offline physics analysis has become standard practice, e.g., suppression of combinatorial background in the LHCb experiment measurement of the $B_s^0-\bar{B}_s^0$ oscillation frequency~\cite{delta-ms}. Classifiers can also aid tasks such as particle identification within event reconstruction, e.g., the CMS boosted event shape tagger, a NN trained to discriminate between possible $t$, $W^\pm$, $Z^0$ and $H$ candidates within an event~\cite{CMS-best}. ML classifiers also find many decision-making applications in industry, for example in the detection of malicious communications~\cite{classifier-phishing}.

ML techniques can also be applied to pattern recognition and anomaly detection—tasks which cannot otherwise be realistically performed on the scales of data presently being analysed. In industry, this is applied intuitively to fraud detection, wherein even subtle changes in transaction  data can indicate fraudulent activity~\cite{fraud-detection}. In HEP, such anomaly detection can be applied to searches, for example the ATLAS search for resonant decays to a Higgs boson~\cite{anomaly-hep}.

\subsection{Hybrid computing architectures}
\label{hybrid-architectures}
\begin{figure*}[h!]
    \centering
    \includegraphics[width=\linewidth,clip]{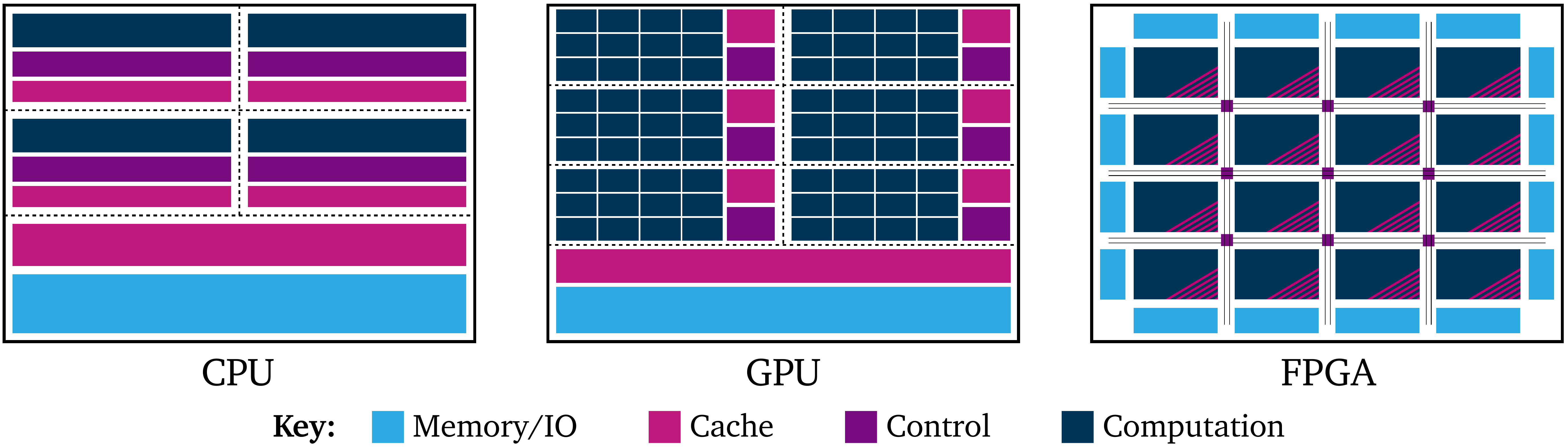}
    \caption{Comparison of CPU, GPU and FPGA architectures, illustrated as schematic diagrams. GPUs typically contain a greater proportion of computational resources than CPUs, with these resources subdivided within each multiprocessor to provide better parallel computing performance. FPGAs take a different approach, comprising many control blocks connected to memory/IO interface and to one another via switches~\cite{architectures}.}
    \label{architectures}
\end{figure*}

Central Processing Units (CPUs) have been used ubiquitously across many fields of research including HEP. CPUs are designed for general purpose computation (i.e., a single processor capable of performing a wide range of tasks with significant variety in computing/memory requirements), with large on-board memory and often multiple processing cores. However, recent advancements in computing have led to the development of accelerators, such as Graphical Processing Units (GPUs), Field-Programmable Gate Arrays (FPGAs), Application-Specific Integrated Chips (ASICs), etc. Accelerators are designed to perform specific tasks significantly faster than a CPU, by virtue of their structural design as represented in Figure~\ref{architectures}. A hybrid computing architecture is thereby a system formed of $2$ or more of the categories of processing units discussed: by designing the computing architecture around the nature of the tasks to be computed, computation can be significantly accelerated~\cite{architectures}.

GPUs contain similar centralized resources to CPUs, but consist of many multiprocessors, each containing a greater proportion of computational resources than an equivalent CPU core. GPUs are thus well-suited to perform computationally intensive tasks in parallel, for example in event reconstruction at LHCb, where tasks such as track reconstruction and particle identification must be completed for every event~\cite{vomBruch-gpus}.

FPGAs are structured very differently, with memory and IO interface connected to many interlinked control blocks formed of simpler logic gate arrangements, often accompanied by a small cache. FGPAs are thus unable to complete complex tasks, though provide significant acceleration of simpler, highly parallelisable tasks. The programmable nature of FPGAs allows for their configuration with ML algorithms for fast performance~\cite{duarte-fpgas, muons-fpgas}.

\section{Early stage researchers}
\label{esrs}
\begin{figure*}[h!]
    \centering
    \includegraphics[width=\linewidth]{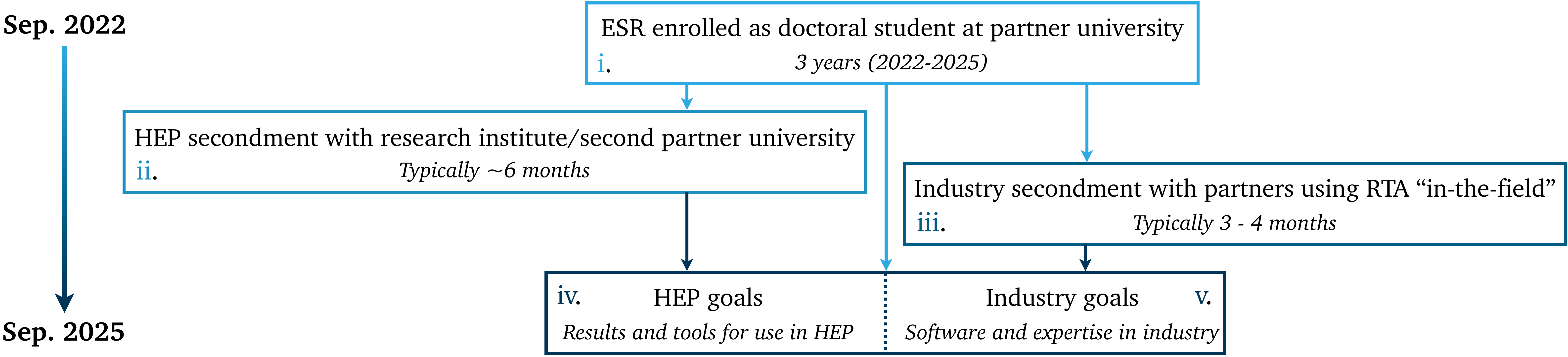}
    \caption{Structure of a SMARTHEP ESR position. Each ESR is enrolled (i.), during which they will undertake secondments with network partners in HEP (ii.) and industry (iii.). Through the combination of primary and secondment work, each ESR will complete goals in HEP (iv.) and industry (v.), discussed in further detail in Section~\ref{outcomes}. Precise durations of the secondments vary between ESR positions.}
    \label{esr-diagram}
\end{figure*}

12 ESRs form the core of the network, with the training and partnerships providing a scaffold for the completion of their respective outcomes. Each ESR is enrolled as a doctoral student at a partner university for 3 years, during which they complete secondments in HEP and industry, as illustrated in Figure~\ref{esr-diagram}. The university of enrolment is generally also the hosting institution, where the majority of the doctoral study is completed (industry-centred ESR positions are enrolled at a university near to the respective industry partner). A secondment in HEP is undertaken either at another partner university or a partner research institute. Each ESR (with the exception of those in industry-centred positions) also undertakes a secondment in industry, working on an RTA project relevant to their research with an industry partner.

To illustrate further the structure of a SMARTHEP ESR position, examples of academia- and industry-centred positions are given below. Details of the positions are given alongside their corresponding label in Figure~\ref{esr-diagram}.

Firstly, an example of an academia-centred position is the ESR based at the University of Heidelberg (i.). Their primary focus is of real-time dark photon searches at LHCb (iv.), furthered by collaboration with the University of Milano-Bicocca (ii.). An industry secondment (iii.) with Verizon Connect is planned, applying ML expertise to the real-time processing of vehicle data (v.).

An industry-centred position is well-typified by the ESR working at IBM, enrolled at Sorbonne University (i.). Their research applies real-time rule induction to fraud detection using real-time rule induction at IBM (v.). An industry secondment (iii.) is not undertaken, since the primary research is undertaken with an industry partner; however, an academic secondment (ii.) is planned with the CNRS research institute, applying similar pattern-recognition techniques to the classification of HEP observations (iv).

\section{Network outcomes}
\label{outcomes}
The network is defined around a set of concrete outcomes, guiding the progression of the network and its participants. These intended outcomes are summarised briefly below.

The core outcomes of the network are a set of goals, completed on behalf of the network partners by the network participants, in particular by the ESRs, who each contribute to at least one whitepaper.  These goals include experiment commissioning, HEP measurements and industrial results. These goals are typically defined with respect to a specific ESR position, e.g., the goal of ``calibration of ALICE TPC for heavy-ion physics''.

Participants will write a series of three whitepapers reviewing the current RTA state-of-the-art, expected to be submitted in late 2023. These whitepapers will place a particular emphasis on the ongoing contributions of the network to their respective topics. The first whitepaper reviews ML applications to HEP in RTA contexts and their corresponding best practices. Such applications range from data-taking, e.g., particle identification algorithms at ALICE, to offline analysis, e.g., anomaly detection for dijet resonance searches at ATLAS~\cite{ALICE-PID, ATLAS-dijet}. The use of hybrid computing architectures by LHC experiments is reviewed in the second whitepaper. Tasks such as selection and reconstruction were significantly accelerated during Run 2 of the LHC (2015-2018). Run 3 deployments of hybrid architectures capitalise on this progress, e.g., in the use of FPGAs in the ALICE Central Trigger System and the use of GPUs in the CMS High Level Trigger farm~\cite{ALICE-CTS, CMS-HLT-farm}. In the third whitepaper, TDAQ systems of LHC experiments are reviewed, with a focus upon best practices for both TDAQ hardware and software. As such, this whitepaper reviews topics ranging from upgrades to the ATLAS TDAQ system, to the Allen framework of the LHCb experiment enabling software trigger operation at a 30~MHz readout rate~\cite{ATLAS-TDAQ, LHCb-Allen}.

As a key network objective, participants are also encouraged to partake in a broad range of training activities, with network resources dedicated to facilitating this. Such activities include the attendance of external specialist industrial training sessions and academic schools, in addition to internally organised workshops and schools (e.g., in Section~\ref{events}).

The work of the ESRs will generate digital assets (e.g., software packages, data processing tools), with many being applicable beyond narrow academic/industrial applications. The network is therefore committed to making any such digital assets Findable, Accessible, Interoperable, and Reusable (FAIR)~\cite{FAIR-principles}. To implement these commitments, a project on GitHub has been created to host such assets \cite{SMARTHEP-github}. Other assets and resources will be made available on the network website \cite{SMARTHEP-website}.

\section{Network events}
\label{events}
Network events provide participants with unique opportunities to meet, exchange ideas and develop. Whilst these events typically cater to the ESR audience, some events have been made available to additional interested early career researchers working/studying at SMARTHEP institutes. 

To commence ESR participation in the network, a kick-off meeting was held at the University of Manchester in November 2022 to formally introduce the ESRs to the network and discuss network objectives and organisation. Amongst the activities of the kick-off meeting were a visit to the Jodrell Bank Centre for Astrophysics and a review paper-writing course by Scriptoria to train ESRs ahead of writing the whitepapers.

In January of 2023, the First SMARTHEP School on Collider Physics and Machine Learning was hosted by the University of Geneva. The school provided a varied programme, including lectures on experimental physics at collider experiments by Anna Sfyrla and on theoretical physics \& Monte Carlo event generators by Torbj\"orn Sj\"ostrand. Additionally, hands-on lessons in machine learning were led by Maurizio Pierini, with seminars also given on multimessenger astronomy and the CERN experimental programme by Teresa Montaruli and Jamie Boyd respectively.

Later events of the network aim to guide the governance of the network, develop ESR expertise and deepen collaborations between participants. The network assembly serves as an annual forum in which to make significant decisions on network policy and governance. Technical hackathons, foreseen for autumn $2023$, give ESRs the opportunity to learn state-of-the-art techniques through their direct application. Hackathons also act as effective team-building activities, promoting problem-solving, rapid prototyping and project management skills, which ESRs can then apply to their own research projects. Accelerator and ML bootcamps, proposed for summer $2024$, will formalise the practical experience of ESRs. Collaboration with industry will culminate in an industry applications school late in the network, by which time it is expected that all ESR industry secondments will have been completed.

\section{Conclusion}
\label{conclusion}
Through the activities of the network outlined in this contribution, SMARTHEP aims to provide valuable contributions to HEP, industry, and the wider community. By the close of the network in September 2025, the 12 ESRs will be well-equipped to progress their research and their careers. The network will play a key role in the commissioning and operation of major LHC experiments and seek to strengthen experimental RTA portfolios by a two-way sharing of knowledge and expertise with industry partners. Such collaboration between academic and industrial network participants will deepen synergies across the field in the application of RTA.

\section*{Acknowledgements}
We acknowledge funding from the European Union Horizon 2020 research and innovation programme, call H2020-MSCA- ITN-2020, under Grant Agreement n. 956086.

\bibliography{references}
\bibliographystyle{JHEP}

\end{document}